# Quasi-Spectral Sparse Bi-Global Stability Analysis of Compressible Channel Flow over Complex Impedance[*]

Iman Rahbari[†] and Carlo Scalo[‡]

*Purdue University, West Lafayette, IN, 47906, United States*

We have developed a fully sparse, compact-scheme based biglobal stability analysis numerical solver applied, for the scope of the current paper, to the investigation of the effects of impedance boundary conditions (IBCs) on the structure of a fully developed compressible turbulent channel flow. A sixth-order compact finite difference scheme is used to discretize the linearized Navier-Stokes equations leading to a Generalized Eigenvalue Problem (GEVP). Sparsity is retained by explicitly introducing derivatives of the perturbation as additional unknowns, increasing the overall problem size (number of columns × number of rows) while significantly reducing the number of non-zeros and the computational cost with respect to traditional implementations yielding otherwise dense matrix blocks. The resulting GEVP is coded in Python and solved employing an Message Passing Interface (MPI) parallelized PETSc-based sparse eigenvalue solver adopting a modified Arnoldi algorithm. Base flow is taken from impermeable isothermal-wall turbulent channel flow simulations at bulk Reynolds number, $Re_b = 6900$ and Mach number, $M_b = 0.85$. The eigenvalue spectrum calculated in the biglobal stability analysis shows distinct groups of modes associated with a discrete set of streamwise wave numbers accommodated by computational domain. An iterative strategy for the imposition of the complex IBCs, which are a nonlinear function of the real-valued (Fourier) frequency in the GEVP, has been devised. The adopted IBC specifically represents an array of sub-surface-mounted Helmholtz cavities with resonant frequency, $f_{res}$, covered by a porous sheet with permeability inversely proportional to the impedance resistance $R$. The tunable resonant frequency has been shown to be an attractor for the instability, yielding a single unstable mode at that frequency. Future work is focused on companion Direct Numerical Simulation (DNS) calculations using a high order compact scheme for the same flow configuration investigating the effects of variable resonant frequency in the streamwise direction $f_{res} = f_{res}(x)$.

## I. Introduction

Porous walls with acoustic absorption characteristics can be modeled via linear acoustic impedance boundary conditions (IBC), which are naturally formulated in the frequency domain as

$$\hat{p}(\Omega) = Z(\Omega)\,\hat{v}_n(\Omega) \tag{1}$$

where $Z(\Omega)$ is a complex-valued function of the angular frequency $\Omega$. $\hat{p}$ and $\hat{v}_n$ are the Fourier transforms of the fluctuating pressure and wall-normal component of the velocity at the boundary (positive if directed away from the fluid side), normalized with the base density and speed of sound (unitary dimensionless base impedance). The current paper will focus on a specific expression of the impedance given by the three-parameter model:

$$Z(\Omega) = R + i\left[\Omega X_{+1} - \Omega^{-1} X_{-1}\right] \tag{2}$$

where $R$ is the impedance resistance and $X_{+1}$ and $X_{-1}$ are the acoustic mass and stiffness, respectively.

---

Our final goal is to develop a global stability analysis framework for prediction of flow response over complex impedance boundary condition of type (2) with parameters variable in streamwise direction and the current paper is the first step in that direction. This effort was inspired by the hydro-acoustic instability observed by Scalo, Bodart, and Lele,[1] who performed Large-Eddy Simulations of compressible channel flow over a complex Impedance Boundary Condition.

## I.A. Background

The effects of porous walls on shear flows has been a topic of formidable research effort, especially in the low-Mach-number limit. Lekoudis[2] performed 1D-LSA of an incompressible boundary layer over two types of permeable boundaries: one modeling a perforated surface over a large chamber (with stabilizing effects); the second one, modeling pores over independent cavities (with negligible effects). Jiménez et al.[3] performed channel flow simulations over active and passive porous walls observing the creation of large spanwise-coherent Kelvin-Helmholtz in the outer layer responsible for frictional drag increase; the near-wall turbulence production cycle was not found to be significantly altered.[4] More recently, Tilton and Cortelezzi[5] investigated channel flow coated with finite-thickness homogeneous porous slabs. Two new unstable modes were discovered, one symmetric and one anti-symmetric, originating from the left branch of eigenvalue spectrum.

We aim to investigate the flow behavior over porous wall in compressible regime. To this end, we first performed linear stability analysis of compressible turbulent and laminar channel flow over purely real impedance in a range of subsonic to transonic Mach numbers.[6] For sufficiently high wall permeability, two unstable modes show up: one representing a bulk pressure mode and another is a standing-wave like mode. They both generate additional Reynolds shear stresses concentrated in the viscous sublayer region. In the current research, we take one step forward, by considering the complex form of Impedance Boundary Condition. We also plan to provide a generalized framework for optimization of the parameters involved in this type of boundary condition, including the optimized distribution of resonant frequency and resistance in streamwise and spanwise directions. The later is not possible without Global Stability Analysis as homogenous assumption of perturbation in these directions will not hold.

The present work focuses on building a biglobal stability analysis framework based upon the previous 1D-LST efforts. This approach has been historically used to study short-wavelength elliptic instability in inviscid vortex flows by Pierrehumbert.[7] Tatsumi and Yoshimura[8] studied the stability of an incompressible laminar flow in a rectangular duct using biglobal stability analysis employing a series of Legendre polynomials. They found a critical value for aspect ratio of the duct above which flow becomes unstable under certain conditions.

Biglobal stability analysis is becoming a widely used tool to study compressible flows. Theofilis and Colonius[9] investigated the stability properties of compressible flow over open cavities using this method. They employed staggered spectral discretization and solved the resulting eigenvalue problem using full QZ algorithm.

Akiki and co-workers[10] employed the biglobal LSA to investigate the compressible flow over walls with injection as a model of rocket motor. They studied viscoacoustic modes that emerge in the vicinity of pure acoustic ones and are more pronounced near the wall.

There are numerous papers in the literature on the use of efficient numerical algorithms to solve the eigenvalue problems. A complete review on the history of using biglobal instability analysis can be found in Theofilis.[11,12] Ehrenstein[13] studied biglobal stability analysis of incompressible channel flow over streamwise riblets. It is one of the earliest works that employs Arnoldi algorithm along with a shift invert technique for efficient calculation of eigenvalues instead of using the full QZ algorithm. Rodríguez and Theofilis[14] studied the parallel performance of their biglobal stability solver which employs a parallel Arnoldi algorithm relying on ScaLAPACK and works with dense matrices. They studied the stability properties of a laminar separation bubble in a flat plate boundary layer on a grid $N_x N_y = 360 \times 64$ on 144 processors in approximately 120 min and massive separation on a NACA0015 airfoil at Reynolds 200 and attack angle of $18°$ on the grid $N_x N_y = 250 \times 250$. The latter case required up to 1 TeraB of RAM and took 22 hours on 1024 processors.

One approach to overcome the computational difficulties arising when dealing with large eigenvalue problems is the use of sparse solvers. A review on the softwares developed for this purpose is available in Hernandez et al.[15] Gennaro and colleagues[16] developed a sparse parallel code to solve the Generalized Eigenvalue Problem (GEVP). Just by using the sparse matrices when working with spectral method, the required memory and CPU time are dropped by around 40% on a $N_x N_y = 50 \times 50$ grid. Several combinations of spectral method, sixth- and fourth-order finite difference schemes are investigated to understand the trade-off between the effect of accuracy and sparsity of the matrices involved in the LSA, e.g. the spectral method gives the highest accuracy but in the least sparse form. They concluded that the best combination, when both directions need the same spatial resolution, is to use a sixth order finite difference scheme in a sparse matrix formation.

There are some numerical algorithms implemented in the context of global stability analysis to achieve the high-order accuracy with a sparse structure including Dispersion-Relation-Preserving (DRP)[17] and FD-q,[18] however, these researches compare their results with the dense formulation of the compact finite difference scheme. The main focus of the current research is to take all of the advantages of compact finite difference scheme and make it possible to use with an affordable cost for global stability analysis purposes.

The present work aims at developing a truly sparse quasi-spectral framework to investigate the Global Stability Analysis of compressible channel flow over impedance walls. A new implementation of compact scheme is proposed in section III.B in order to provide the spectral-like accuracy with sparse matrices. Iterative schemes employed for solving the GEVP, i.e., a modified version of Arnoldi algorithm along with shift-invert technique, are discussed in section III.C. Computational performance of the new sparse implementation is compared against the traditional one in section III.E. Section III.F presents an iterative approach in order to implement IBC in the context of Linear Stability Analysis. Results of global stability analysis for the physical problem of interest are shown in section IV.

## II. Linearized Governing Flow Equations

A generic instantaneous quantity, $a(x,y,z,t)$, is decomposed into a base state, $\mathcal{A}(x,y,z)$, and a three-dimensional fluctuation, $a'(x,y,z,t)$. The governing equations are then linearized assuming ideal gas law and retaining only the first order fluctuations. The resulting set of equations for biglobal stability analysis is similar to the one reported in Theofilis and Colonius.[9] In the present study, fluctuations are in the form:

$$a'(x,y,z,t) = \hat{a}(x,y)e^{i(\beta z - \omega t)} \tag{3}$$

where $\omega$ is the complex frequency of each mode, $\Re\{\omega\}$ the phase speed and $\Im\{\omega\}$ the growth rate of each mode.

The dimensional governing equations for conservation of mass, momentum, energy as well as equation of state are written in the following:

$$\frac{\partial \rho'}{\partial t} + \frac{\partial}{\partial x}\left(\bar{\rho}u'\right) + \frac{\partial}{\partial x}\left(\rho'\bar{u}\right) + \frac{\partial}{\partial y}\left(\bar{\rho}v'\right) + \frac{\partial}{\partial y}\left(\rho'\bar{v}\right) = 0 \tag{4}$$

$$\bar{\rho}\left(\frac{\partial u'}{\partial t} + u'\frac{\partial \bar{u}}{\partial x} + \bar{u}\frac{\partial u'}{\partial x} + v'\frac{\partial \bar{u}}{\partial y} + \bar{v}\frac{\partial u'}{\partial y}\right) + \rho'\left(\bar{u}\frac{\partial \bar{u}}{\partial x} + \bar{v}\frac{\partial \bar{u}}{\partial y}\right) = \tag{5}$$
$$\frac{\partial p'}{\partial x} + \frac{\partial}{\partial x}\left[l_2\bar{\mu}\frac{\partial u'}{\partial x} + l_2\mu'\frac{\partial \bar{u}}{\partial x} + l_0\bar{\mu}\frac{\partial v'}{\partial y} + l_0\mu'\frac{\partial \bar{v}}{\partial y}\right] + \frac{\partial}{\partial y}\left[\bar{\mu}\left(\frac{\partial u'}{\partial y} + \frac{\partial v'}{\partial x}\right) + \mu'\left(\frac{\partial \bar{u}}{\partial y} + \frac{\partial \bar{v}}{\partial x}\right)\right]$$

$$\bar{\rho}\left(\frac{\partial v'}{\partial t} + u'\frac{\partial \bar{v}}{\partial x} + \bar{u}\frac{\partial v'}{\partial x} + v'\frac{\partial \bar{v}}{\partial y} + \bar{v}\frac{\partial v'}{\partial y}\right) + \rho'\left(\bar{u}\frac{\partial \bar{v}}{\partial x} + \bar{v}\frac{\partial \bar{v}}{\partial y}\right) = \tag{6}$$
$$\frac{\partial p'}{\partial y} + \frac{\partial}{\partial y}\left[l_2\bar{\mu}\frac{\partial v'}{\partial y} + l_2\mu'\frac{\partial \bar{v}}{\partial y} + l_0\bar{\mu}\frac{\partial u'}{\partial x} + l_0\mu'\frac{\partial \bar{u}}{\partial x}\right] + \frac{\partial}{\partial x}\left[\bar{\mu}\left(\frac{\partial u'}{\partial y} + \frac{\partial v'}{\partial x}\right) + \mu'\left(\frac{\partial \bar{u}}{\partial y} + \frac{\partial \bar{v}}{\partial x}\right)\right]$$

$$\rho' = \frac{P'}{R_g T'} - \frac{\bar{\rho}}{\bar{T}}T' \tag{7}$$

$$\bar{\rho}C_p \left( \frac{\partial T'}{\partial t} + \bar{u}\frac{\partial T'}{\partial x} + u'\frac{\partial \overline{T}}{\partial x} + \bar{v}\frac{\partial T'}{\partial y} + v'\frac{\partial \overline{T}}{\partial y} \right) + \rho'C_p \left( \bar{u}\frac{\partial \overline{T}}{\partial x} + \bar{v}\frac{\partial \overline{T}}{\partial y} \right) = \qquad (8)$$

$$\frac{\partial \bar{k}}{\partial x}\frac{\partial T'}{\partial x} + \bar{k}\frac{\partial^2 T'}{\partial x^2} + \frac{\partial k'}{\partial x}\frac{\partial \overline{T}}{\partial x} + k'\frac{\partial^2 \overline{T}}{\partial x^2} + \bar{k}\frac{\partial^2 T'}{\partial y^2} + \frac{\partial k'}{\partial y}\frac{\partial \overline{T}}{\partial y} + k'\frac{\partial^2 \overline{T}}{\partial y^2} + \frac{\partial p'}{\partial t} + \bar{u}\frac{\partial p'}{\partial x} + u'\frac{\partial \bar{p}}{\partial x} + \bar{v}\frac{\partial p'}{\partial y} +$$

$$v'\frac{\partial \bar{p}}{\partial y} + 2l_2\bar{\mu}\left[\left(\frac{\partial \bar{u}}{\partial x}\frac{\partial u'}{\partial x} + \frac{\partial \bar{v}}{\partial y}\frac{\partial v'}{\partial y}\right)\right] + \bar{\mu}\left[2l_0\left(\frac{\partial \bar{u}}{\partial x}\frac{\partial v'}{\partial y}\right) + 2l_0\left(\frac{\partial \bar{v}}{\partial y}\frac{\partial u'}{\partial x}\right) + 2\left(\frac{\partial u'}{\partial y} + \frac{\partial v'}{\partial x}\right)\left(\frac{\partial \bar{u}}{\partial y} + \frac{\partial \bar{v}}{\partial x}\right)\right] +$$

$$l_2\mu'\left[\left(\frac{\partial \bar{u}}{\partial x}\right)^2 + \left(\frac{\partial \bar{v}}{\partial y}\right)^2\right] + \mu'\left[\left(\frac{\partial \bar{u}}{\partial y} + \frac{\partial \bar{v}}{\partial x}\right)^2 + 2l_0\left(\frac{\partial \bar{u}}{\partial x}\frac{\partial \bar{v}}{\partial y}\right)\right]$$

where $x$, $y$, and $z$ are, respectively, the streamwise, wall-normal and spanwise coordinates. Quantities shown by $\overline{()}$ are taken from the base flow calculations while $()'$ variables denote the fluctuating terms. Having this set of equations, one has a wide range of choices for non-dimensionalization. In this paper, all quantities, are normalized with the channel's half-width, the bulk density (constant for channel flow), and the speed of sound, temperature and dynamic viscosity at the wall. Following this non-dimensionalization, the gas constant ($R_g$), becomes equal to $1/\gamma$ leading to the equation of state $p = \gamma^{-1}\rho T$. $Re_a$ is the Reynolds number based on the channel's half-width and speed of sound at the wall temperature, which is related to the bulk Mach and Reynolds numbers via $Re_b = M_b\,Re_a$. Variation of viscosity with temperature is modeled using power-law, $\mu = T^n$, where $n = 0.70$ is the viscosity exponent. Prandtl number, $Pr$, is 0.72 throughout the entire paper.

## III. Numerical Setup for Global Linear Stability Analysis

### III.A. Discretization scheme

Compact schemes are a class of implicit finite difference numerical discretization methods that are capable of retaining the spectral-like accuracy using a reduced stencil. Numerical approximation to the first derivative using such scheme reads:

$$\alpha f'_{i-1} + f'_i + \alpha f'_{i+1} = \frac{a}{2h}(f_{i+1} - f_{i-1}) + \frac{b}{4h}(f_{i+2} - f_{i-2}) \qquad (9)$$

$$a = \frac{2}{3}(2+\alpha), \quad b = \frac{1}{3}(-1+4\alpha)$$

Setting $\alpha = \frac{1}{4}$ gives the fourth order truncation error, however, $\alpha = \frac{1}{3}$ provides a sixth order scheme. Results in this paper are shown using the later value of $\alpha$. In matrix notation, equation (9) reads:

$$L\,\mathbf{f}' = R\,\mathbf{f} \qquad (10)$$

Where $L$ and $R$ are banded matrices. The discrete derivative matrix operator calculated explicitly using this scheme

$$D = L^{-1}R \qquad (11)$$

is dense (figure 1), therefore, results in a computational cost very similar to spectral methods. In the linear stability calculations, the only advantage of using compact scheme following this formulation over Fourier-based spectral methods is the ease of working with generalized curvilinear coordinate system to solve the linearized Navier-Stokes equations in mildly complex geometries.

In Global Stability Analysis, it is more convenient to have spatial derivatives explicitly in terms of primitive variables so that the unknown vector only contains the primitive variables. In the literature[17] and,[18] whenever compact scheme is employed to discretize linearized Navier-Stokes equations, $D$ matrix is used to find the first and second derivatives, $\mathbf{f}''$ and $\mathbf{f}'$, explicitly in terms of $\mathbf{f}$. The optimum approach to use compact is to incorporate $L$ and $R$ matrices inside the system of equation without any inversion to keep the system as sparse as possible. The new variable arrangement proposed in this research is designed to make this accomplished.

Discretized linear stability equations can be arranged in the form of a generalized eigenvalue problem:

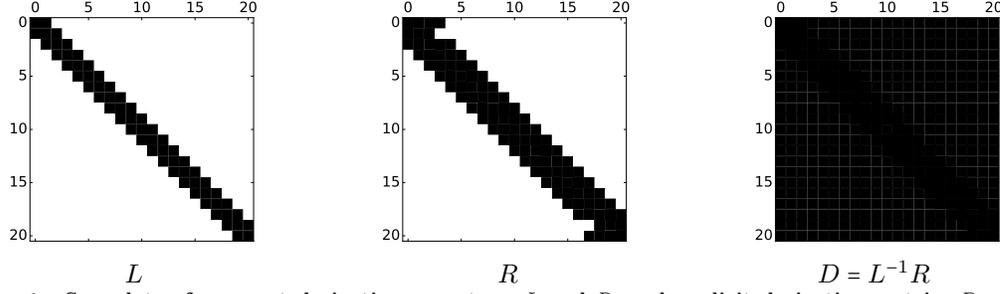

Figure 1. Spy plots of compact derivative operators, $L$ and $R$, and explicit derivative matrix, $D$. $L$ and $R$ operators are banded (tri- and penta-diagonal) matrices while the $D$ matrix is dense.

$$[A]\Psi = \Lambda[B]\Psi \tag{12}$$

or

$$\begin{pmatrix} A_{u1} & A_{u2} & A_{u3} & A_{u4} & A_{u5} \\ A_{v1} & A_{v2} & A_{v3} & A_{v4} & A_{v5} \\ A_{w1} & A_{w2} & A_{w3} & A_{w4} & A_{w5} \\ A_{p1} & A_{p2} & A_{p3} & A_{p4} & A_{p5} \\ A_{t1} & A_{t2} & A_{t3} & A_{t4} & A_{t5} \end{pmatrix} \begin{pmatrix} \hat{\mathbf{u}} \\ \hat{\mathbf{v}} \\ \hat{\mathbf{w}} \\ \hat{\mathbf{p}} \\ \hat{\mathbf{T}} \end{pmatrix} = \omega \begin{pmatrix} B_{u1} & B_{u2} & B_{u3} & B_{u4} & B_{u5} \\ B_{v1} & B_{v2} & B_{v3} & B_{v4} & B_{v5} \\ B_{w1} & B_{w2} & B_{w3} & B_{w4} & B_{w5} \\ B_{p1} & B_{p2} & B_{p3} & B_{p4} & B_{p5} \\ B_{t1} & B_{t2} & B_{t3} & B_{t4} & B_{t5} \end{pmatrix} \begin{pmatrix} \hat{\mathbf{u}} \\ \hat{\mathbf{v}} \\ \hat{\mathbf{w}} \\ \hat{\mathbf{p}} \\ \hat{\mathbf{T}} \end{pmatrix} \tag{13}$$

Various blocks in $[A]$ and $[B]$ correspond to the linearized Navier-Stokes formulation in section II. Most of the papers in the literature use explicit derivative matrices, either using spectral method, compact scheme or explicit finite difference method, such that all the terms in the linearized stability equations involving derivatives of unknowns can be written as a product of a derivative matrix and that unknown, e.g. $D_{xy}\hat{u} = D_x D_y \hat{u}$. Therefore:

$$\hat{\mathbf{u}} = [\hat{u}_0, \hat{u}_1, \cdots \hat{u}_{n-1}], \quad \hat{\mathbf{v}} = [\hat{v}_0, \hat{v}_1, \cdots \hat{v}_{n-1}], \quad \hat{\mathbf{w}} = [\hat{w}_0, \hat{w}_1, \cdots \hat{w}_{n-1}] \\ \hat{\mathbf{p}} = [\hat{p}_0, \hat{p}_1, \cdots \hat{p}_{n-1}], \quad \hat{\mathbf{T}} = [\hat{T}_0, \hat{T}_1, \hat{T}_2, \cdots \hat{T}_{n-1}] \tag{14}$$

where $n = N_x N_y$. `spy` plot of $[A]$ and $[B]$ matrices using explicit implementation of compact scheme are shown in figure 2.

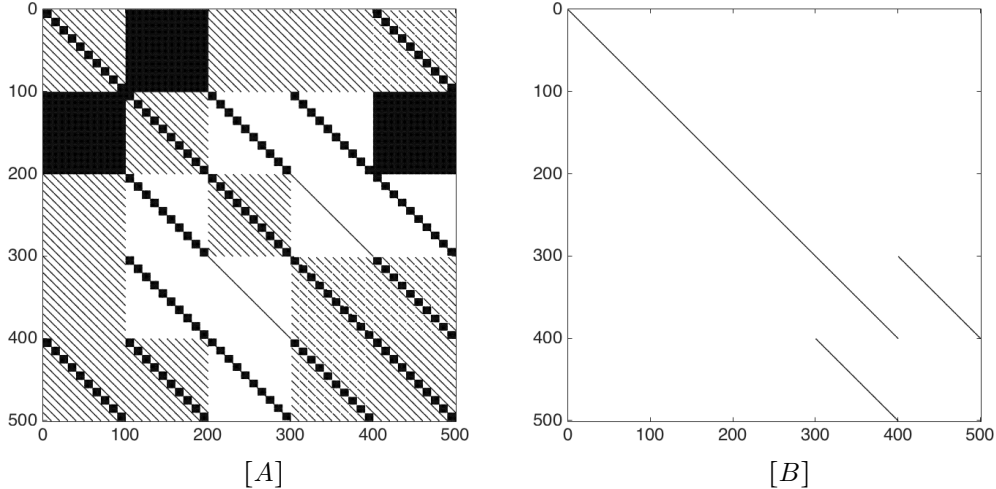

Figure 2. Spy plots of $[A]$ and $[B]$ matrices using dense implementation of compact scheme for a grid of $10 \times 10$.

### III.B. Sparse Implementation of Compact Schemes For Linear Stability Analysis

As explained before, to fully exploit the advantages of compact schemes, one should use the sparse form, incorporating both $L$ and $R$ matrices when setting up the GEVP. For this purpose, the first and second spatial derivatives of the primitive variables are also included in the eigenvector $\Psi$ so that, unlike equation (14), the first block of unknowns, $\hat{\mathbf{u}}$, becomes:

$$\hat{\mathbf{u}} = \left[ \underbrace{\hat{u}_0, \cdots \hat{u}_{n-1}}_{\hat{u}_i}, \underbrace{\hat{u}_{x_0}, \cdots \hat{u}_{x_{n-1}}}_{\hat{u}_{x_i}}, \underbrace{\hat{u}_{xx_0}, \cdots \hat{u}_{xx_{n-1}}}_{\hat{u}_{xx_i}}, \underbrace{\hat{u}_{xy_0}, \cdots \hat{u}_{xy_{n-1}}}_{\hat{u}_{xy_i}}, \underbrace{\hat{u}_{y_0}, \cdots \hat{u}_{y_{n-1}}}_{\hat{u}_{y_i}}, \underbrace{\hat{u}_{yy_0}, \cdots \hat{u}_{yy_{n-1}}}_{\hat{u}_{yy_i}} \right]^T \quad (15)$$

Similarly, one can form the unknown blocks for $\hat{\mathbf{v}}$ and $\hat{\mathbf{w}}$. The temperature vector, $\hat{\mathbf{T}}$, will be similar except that cross derivative does not come into the calculation:

$$\hat{\mathbf{T}} = \left[ \underbrace{\hat{T}_0, \cdots \hat{T}_{n-1}}_{\hat{T}_i}, \underbrace{\hat{T}_{x_0}, \cdots \hat{T}_{x_{n-1}}}_{\hat{T}_{x_i}}, \underbrace{\hat{T}_{xx_0}, \cdots \hat{T}_{xx_{n-1}}}_{\hat{T}_{xx_i}}, \underbrace{\hat{T}_{y_0}, \cdots \hat{T}_{y_{n-1}}}_{\hat{T}_{y_i}}, \underbrace{\hat{T}_{yy_0}, \cdots \hat{T}_{yy_{n-1}}}_{\hat{T}_{yy_i}} \right]^T \quad (16)$$

and $\hat{\mathbf{p}}$ will be:

$$\hat{\mathbf{p}} = \left[ \underbrace{\hat{p}_0, \cdots \hat{p}_{n-1}}_{\hat{p}_i}, \underbrace{\hat{p}_{x_0}, \cdots \hat{p}_{x_{n-1}}}_{\hat{p}_{x_i}}, \underbrace{\hat{u}_{p_0}, \cdots \hat{p}_{y_{n-1}}}_{\hat{p}_{y_i}} \right]^T \quad (17)$$

To set up $[A]$ and $[B]$ matrices, the linear stability equations are written at the first row blocks, i.e., the $x$ and $y$-momentum equations are written in the rows associated with $\hat{u}_i$ and $\hat{v}_i$ blocks and compact derivative formula (10) are incorporated in the newly appeared blocks, e.g., rows corresponding to $\hat{u}_{x_i}$, $\hat{u}_{xx_i}$. For better description of this approach, the structure of the first two blocks of $[A]$ matrix, $A_{u11}$ and $A_{u12}$, are shown in figure 3. Finally, $[A]$ and $[B]$ matrices for a $10 \times 10$ grid are shown in figure 4. Comparing figures 2 and

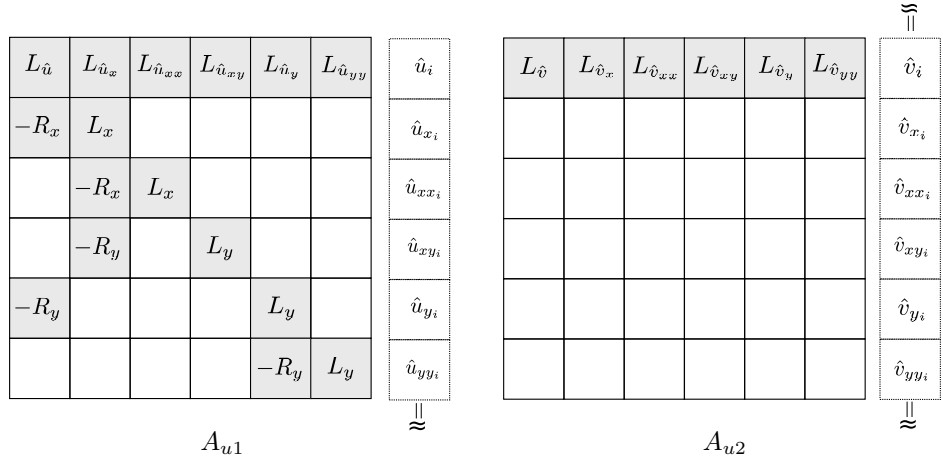

Figure 3. Sample structure of $A_{u1}$ and $A_{u2}$ as part of the $A$ matrix (13). Blocks having non-zero entries are shaded while the ones with all zero elements are left white. The corresponding segments in the array $\Psi$ are shown in the right

4 shows that using the new formulation, the final matrices are much sparser, however bigger in dimensions. The number of non-zero elements in these matrices are shown in table 1. The advantages of this method become more pronounced by increasing the number of grid points so that for a moderate grid size ($128 \times 128$), the new approach produces the matrices that are more than 280 times sparser.

### III.C.  Parallel Sparse Eigenvalue Solver

The generalized eigenvalue problem (12) is solved using SLEPc, Scalable Library for Eigenvalue Problem Computations.[19] The Arnoldi algorithm, based on Krylov subspace iteration, is used to solve the eigenvalue problem. A brief description of this algorithm is included in the following.

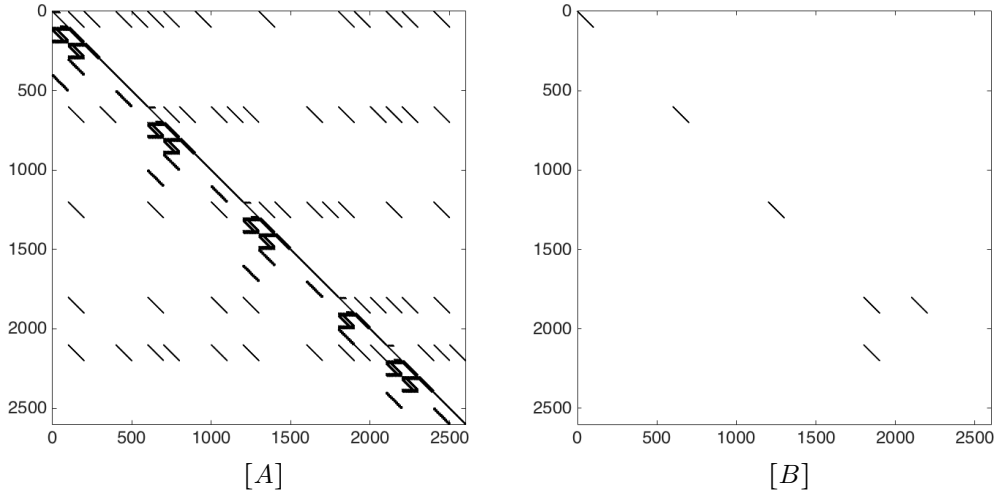

**Figure 4. Spy plots of $[A]$ and $[B]$ matrices using the implicit(sparse) implementation of compact scheme for a grid of $10 \times 10$**

**Table 1. Comparison of number of non-zero elements, nnz(), in matrices $[A]$ and $[B]$ (12), generated using sparse versus dense compact scheme implementation**

| Grid Size $(N_x \times N_y)$ | nnz($A$) Sparse | nnz($A$) Dense | nnz($B$) Sparse | nnz($B$) Dense |
|---|---|---|---|---|
| $16 \times 16$ | 46,680 | 263,073 | 5,610 | 1,530 |
| $32 \times 32$ | 182,807 | 3,684,908 | 22,506 | 6,138 |
| $64 \times 64$ | 740,565 | 54,590,571 | 90,090 | 24,570 |
| $128 \times 128$ | 2,980,309 | 840,071,345 | 360,426 | 98,298 |

### III.C.1. Arnoldi Algorithm

Let matrix $M$ be a generic complex valued matrix for which the decomposition $M = QHQ^*$ holds where $Q$ and $H$ are unitary and upper Hessenberg matrices, respectively. One can write this decomposition for the first $m$ columns as:

$$M_{[n \times n]} Q'_{[n \times m]} = Q'_{[n \times m+1]} H'_{[m+1 \times m]} \tag{18}$$

where prime denotes a portion of the original matrix.

$$\begin{bmatrix} & & \\ & M & \\ & & \end{bmatrix} \begin{bmatrix} q_1 & q_2 & \cdots & q_m \end{bmatrix} = \begin{bmatrix} q_1 & q_2 & \cdots & q_{m+1} \end{bmatrix} \begin{bmatrix} h_{11} & h_{12} & \cdots & h_{1m} \\ h_{21} & h_{22} & \cdots & h_{2m} \\ & \ddots & & \vdots \\ & & \ddots & \vdots \\ & & & h_{m+1,m} \end{bmatrix}$$

The basic Arnoldi algorithm uses the stabilized Gram-Schmidt process to find a sequence of orthonormal vectors, e.g. $q_1, q_2, q_3, \cdots$, such that for every $m$ and starting from a random $b$ vector, $\mathcal{K}_m(M, b) = \text{span}\{q_1, q_2, \cdots q_m\}$. This algorithm can be summarized as:

**Algorithm 1** Basic Arnoldi iteration algorithm

**procedure** BASIC ARNOLDI ITERATION ALGORITHM
    selecting $b$ arbitrarily which gives $q_1 = \frac{b}{||b||}$
    **for** $m = 1, 2, \cdots$ **do**
        $p = M q_m$
        **for** $i = 1, 2, \cdots, m$ **do**
            $h_{im} = q_i^* p$
            $p = p - h_{im} q_i$
        **end for**
        $h_{m+1,m} = ||p||$
        $q_{m+1} = \frac{p}{h_{m+1,m}}$
    **end for**
**end procedure**

Removing the last row of $H'_{m+1 \times m}$ leads to $\tilde{H}_{[m \times m]}$ which depending on $m$ can be much smaller than $M$ matrix. Eigenvalues of $\tilde{H}$ are good approximations of those of $M$ matrix and $q_1$ to $q_m$ are the eigenvectors. The algorithm implemented in this study, through SLEPc, is a variant of the original Arnoldi algorithm called Explicitly Restarted Arnoldi with locking. More information on this can be found on SLEPc user manual.[20]

### III.C.2. *Shift-and-Invert Technique*

The generalized eigenvalue problem (12) can be transformed into a standard eigenvalue problem:

$$\tilde{A} u = \tilde{\Lambda} u \tag{19}$$

where $\tilde{A} = (A - \sigma B)^{-1} B$ and $\tilde{\Lambda} = \frac{1}{\omega - \sigma}$. Using this transformation, the eigenvalues close to the $\sigma$ become the largest in magnitude and will converge quickly in Krylov-based algorithms. All calculations in these algorithms that require $\tilde{A}$ are handled implicitly through PETSc so that all matrix-vector products are viewed as solution of a linear system of equation. At each Krylov iteration, one system of equation should be solved, using direct scheme which is one of the most time consuming parts of the process of finding the eigenvalues. Direct solvers find the LU decomposition of the matrix and then solve for the multiple right hand sides which the later only requires a backward and forward substitutions. The distributed version of the SuperLU package is used as the direct solver which employs Message Passing Interface (MPI) to perform the Gaussian elimination in parallel.[21, 22] Matrices are partitioned so that each processor reads one block-row. Factorized matrices ($L$ and $U$) are also distributed amongst several processors which necessitates performing the backward/forward substitutions in the distributed form that causes difficulties when using many processors.

### III.D. Validation of the Biglobal Solver

In this subsection we aim to study the accuracy of the developed biglobal solver using the new and traditional implementations of the compact scheme. For this purpose, the leading edge boundary layer problem is chosen which is originally studied by Lin and Malik[23] using a spectral solver for an incompressible flow and then is extended to the compressible regime by Theofilis et al.[24] One interesting fact about this test case is the ease of calculating the base flow by solving a set of coupled Ordinary Differential Equations, presented in equations (20), with high accuracy and low computational cost.

$$\begin{aligned}
V' &= -U + V \frac{T'}{T} \\
U'' &= \frac{1}{\mu} \left( \frac{U^2 + VU'}{T} - 1 - \frac{\partial \mu}{\partial T} T' U' \right) \\
W'' &= \frac{1}{\mu} \left( \frac{VW'}{T} - \frac{\partial \mu}{\partial T} T' W' \right) \\
T'' &= \frac{Pr}{\mu} \left( -\frac{\partial \mu}{\partial T} \frac{T'^2}{Pr} + \frac{T'V}{T} - (\gamma - 1) M^2 \mu W'^2 \right)
\end{aligned} \tag{20}$$

This set of ODEs, subjected to the boundary conditions (21), is solved using the fourth-order Runge-Kutta scheme.

$$U(0) = V(0) = W(0) = 0 \quad \text{and} \quad U(\infty) = W(\infty) = T(\infty) = 1 \quad (21)$$

Velocity components as well as temperature base field are calculated after solving the above-mentioned equations resulting in the base flow vector:

$$\mathcal{A}(x,y,z) = (u,v,w,\rho,T) = \left(\frac{xU(y)}{\text{Re}}, \frac{V(y)}{\text{Re}}, W(y), \frac{1}{T(y)}, T(y)\right) \quad (22)$$

Linearized Navier-Stokes equations, described in (5) to (9), are solved using both implementations of compact scheme at Re = 800, M = 0.02 and $\beta$ = 0.255. At the wall, all components of the velocity fluctuations are set to zero $(\hat{u}, \hat{v}, \hat{w}) = 0$ as well as temperature fluctuation $\hat{T} = 0$. Zero Neumann boundary condition is imposed for pressure at the wall, $\partial \hat{p}/\partial y = 0$. At the far-field, all perturbations are assumed to decay to zero. Linear extrapolation is imposed at the left and right boundaries in $x$-direction.

Two consecutive grid transformations are used to map the uniform grid (in $y$ direction denoted by $\chi \in [-1,1]$) to the one used by Lin and Malik[23] in order to achieve a fair comparison. The first one, $\eta = \tanh(\gamma\chi)/\tanh(\chi)$ with $\gamma = 2$, provides a non-uniform grid with points clustered near the boundaries, however, the second transformation places half of the points in a very small distance adjacent to the wall (shown by $y_i$):

$$y = a\frac{1+\eta}{b-\eta} \quad \text{where} \quad a = \frac{y_i y_\infty}{(y_\infty - 2y_i)} \quad \text{and} \quad b = 1 + \frac{2a}{y_\infty} \quad (23)$$

In the current study, $y_i = 0.5$ and $y_\infty = 100$ are considered. In $x$ direction, a uniform grid is considered, $x \in [-100, 100]$. Eigenvalues of the first and second most unstable modes on a 48 × 48 grid are presented in the table 2 which shows the accuracy of the developed code using compact scheme in Linear Stability Analysis.

Table 2. Biglobal stability analysis of leading edge boundary layer at Re = 800, M = 0.02 and $\beta$ = 0.255 on a 48 × 48 grid where $c_i = \omega_i/\beta$ and $c_r = \omega_r/\beta$. Subscripts GH and A1 represent the Görtler-Hämmerlin and first anti-symmetric modes. Deviating digits are underlined.

| Grid: 48 × 48 | $c_{r\text{GH}}$ | $c_{i\text{GH}}$ | $c_{r\text{A1}}$ | $c_{i\text{A1}}$ |
|---|---|---|---|---|
| Lin and Malik[23] | 0.35840982 | 0.00585325 | 0.35791970 | 0.00409887 |
| Theofilis et al.[24] | 0.35844151 | 0.00585646 | 0.35793726 | 0.00401330 |
| Current Study: Sparse | 0.35844071 | 0.00585467 | 0.35795061 | 0.00410000 |
| Current Study: Dense | 0.35844457 | 0.00584620 | 0.35795353 | 0.00409183 |

### III.E. Assessment of the computational performance of the Biglobal Solver

After validating the solver against the available data in the literature, computational performance of both implementations of compact scheme should be assessed. All the calculations are performed on the *Rice* cluster at Purdue University featuring two 10-Core Intel Xeon-E5 with 64 GB of memory per node.

Dense (traditional) implementation demands much more memory such that the eigenvalue computations on the grids finer than 80 × 80 was not possible on our machines. Moreover, this implementation is much problematic to work with in parallel, i.e., calculations on the small grids, up to 30 × 30, can only be performed on less than or equal to 4 processors and calculations on the relatively moderate grids, up to 70 × 70, can be parallelized on up to 8 processors. However, the calculations using the new implementation are more stable providing more freedom in choosing the number of processors.

Total number of non-zero elements after factorization in the $L + U$ matrix is considered as a measure of cost of the implementation. The new approach creates less number of fill-ins, except at very small grid sizes, so that at the 70 × 70 grid, $L + U$ matrix generated using this approach contains half number of non-zero elements of the explicit implementation's. This ratio rapidly increases by increasing the number of grid points causing memory issues that prevents us from performing the calculations on the finer grids using the dense implementation. Total memory highmark is also measured for both approaches showing that the new implementation can significantly reduce the memory requirement. A guid for eyes, following the trend of the finest possible grids for the dense implementation, is plotted to give an idea about the cost of using different implementations on finer grids.

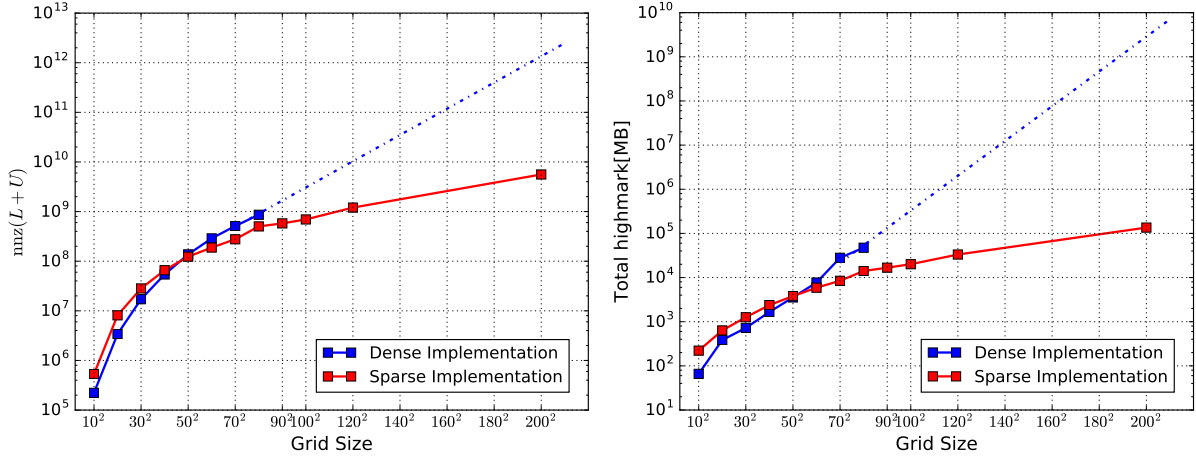

Figure 5. Total number of non-zero elements (left) and total highmark (right) for sparse and dense implementations of compact scheme

### III.F.  Implementation of Complex Impedance Boundary Condition

To impose the Impedance Boundary Condition, equation (1) must be written at the boundary. Note that in that equation, $\Omega$ refers to the angular frequency, and, as mentioned earlier, $\mathfrak{R}(\omega)$ in the fluctuation form (3), represents the mode frequency. Imposing this boundary condition leads to a non-linear generalized eigenvalue problem where specific rows in the matrices depend on the real part of the eigenvalue. To solve this problem, an iterative approach is proposed based on the physical properties of this boundary condition. This algorithm is presented in form of a pseudo code in the following:

---
**Algorithm 2** Solving the Non-linear GEVP
---
1: **procedure** SOLVING THE NON-LINEAR GEVP
2:     set $Z(\Omega) = R$
3:     solve the eigenvalue problem:
$$Av_i = \Omega_i B v_i$$
4:     start tracking the eigenvalues in the region of interest:
5:     **for** $i = 1, nEV$ **do**
6:         set $\Omega_c = \Omega_i$
7:         **for** $j = 1, maxIter$ **do**
8:             find the new value for $Z(\Omega)$: $Z(\Omega) = R - i\left[X_{(-1)}\mathfrak{R}(\omega_c)^{-1} - X_{(+1)}\mathfrak{R}(\omega_c)\right]$
9:             solve the eigenvalue problem $Av_j = \omega_j B v_j$:    ▷ use shift-invert algorithm targeted at $\omega_i$ seeking only 1 mode
10:           **if** $err = \|\omega_j - \omega_c\|_2 < tol$ **then**
11:               break
12:           **end if**
13:           set $\omega_c = \omega_i$
14:         **end for**
15:     **end for**
16: **end procedure**
---

## IV.  Stability Analysis of Compressible Channel Flow over IBC

Computational setup considered in this study is a fully-developed turbulent channel flow with porous bottom and hard top walls. Figure 6 shows a schematic view of this flow configuration. No-slip boundary condition for streamwise and spanwise velocity components is imposed at both walls. Impedance boundary condition is written at the bottom wall while wall-normal velocity is set to zero at the top wall to represent the hard wall boundary condition and both walls are assumed adiabatic. Calculations are performed for bulk Mach number $M_b = 0.85$ and bulk Reynolds number $Re_b = 6900$. Two following parameters, undamped resonant angular frequency $\omega_{res}$, and damping ratio $\zeta$, should be defined in order to have a better intuition

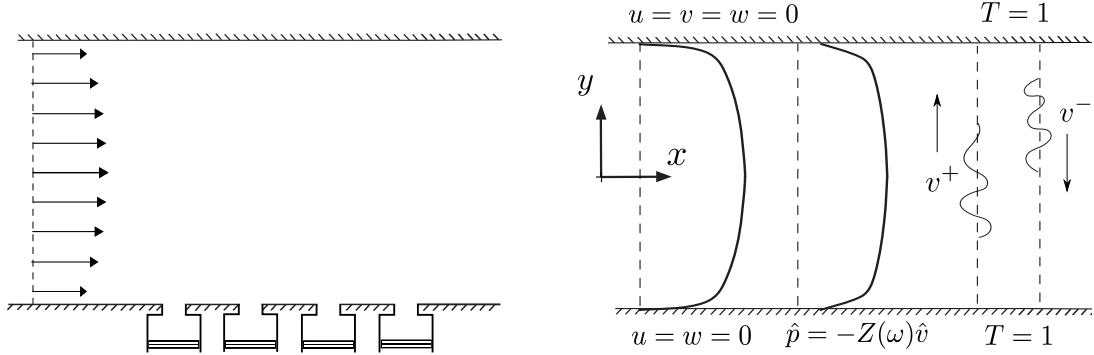

Figure 6. On the left, computational setup for stability analysis of compressible channel flow over (one) impedance wall. On the right, an illustration of the physical equivalent of the investigated computation setup

about the IBC:

$$w_{res} = \sqrt{\frac{X_{-1}}{X_{+1}}}, \qquad \zeta = \frac{1+R}{2\omega_{res}X_{+1}} \tag{24}$$

Scalo et al.[1] concluded that $\zeta$ has an insignificant effect on the simulation results, so, here we consider $\zeta = 0.5$. In this research, $R$ is fixed to $R = 1.0$ and we only focus on the effect of resonant frequency on the flow response.

The basic flow for the following analysis is the Reynolds-averaged turbulent flow taken from high-fidelity impermeable isothermal wall Direct Numerical Simulations carried out with a parallel compact-finite-difference Navier-Stokes solver[25] (CFDSU). The solver adopts a staggered variable arrangement for the conservative variables to improve the quality of the numerical scheme.[26] Time advancement is carried out employing a third-order Runge-Kutta method.

### IV.A. Base Flow Calculation

The computational domain considered for the turbulent simulations, used afterward to calculated the base flow variables, is $L_x \times L_y \times L_z = 8 \times 2 \times 4$ which is discretized with a grid of size $N_x \times N_y \times N_z = 256 \times 128 \times 192$ resulting in a quasi-DNS resolution of $\Delta x^+ \sim 13.22$, $\Delta z^+ \sim 8.81$, $\Delta y^+_{min} = 0.47$. The superscript $^+$ indicates classic wall-units, $\delta_\nu = u_\tau \rho_w / \mu_w$, where $u_\tau = \sqrt{\tau_w/\rho_w}$ is the friction velocity and $\rho_w$ and $\mu_w$ are the density and dynamic viscosity evaluated at the wall and the wall-shear stress is $\tau_w = \mu_w \, \partial \overline{u}/\partial y|_w$ where $\overline{(\ )}$ indicates Reynolds averaging. Friction Reynolds number, $\text{Re}_\tau = u_\tau \delta/\nu = 417.4$.

### IV.B. Results

Generalized eigenvalue problem (12) with aforementioned boundary conditions is solved on $128 \times 128$ grid points, for which in $y$-direction, grid is non-uniformly distributed following $y = \tanh(\gamma\eta)/\tanh(\eta)$, where $\eta$ is evenly spaced in $y$- direction and stretching factor $\gamma$ is 1.0.

In our previous work, we have shown that decreasing the Resistance of impedance destabilizes the flow.[6] It is also reported in Scalo et al.[1] that damping ratio $\zeta$ does not change the flow response, significantly. In this research, we only focus on the effect of resonant frequency on the flow behavior by keeping the Resistance and damping ratio fixed to $R = 1$ and $\zeta = 0.5$, respectively. Resonant frequency on the bottom wall is kept constant in $x$-direction at each test case and is selected from the set $\omega_{res} = [0.596, 1.046, 1.443, 3.779]$ spanning approximately one decade of frequencies. Eigenvalue spectrum for this set of numerical experiments are reported in figure 7. It is observed that the tunable resonant frequency acts as an attractor for the instability, yielding one single unstable mode at each resonant frequency. Modes whose frequencies depart from the resonant frequency are less affected by the imposition of IBC. Pressure and wall-normal velocity eigenfunctions associated with the unstable modes are plotted in figures 8 and 9 for the aforementioned range of resonant frequencies. It is observed that the structure of the unstable modes is consistent with Kelvin-Helmholtz-like rollers that are confined near the wall, minimally affecting the outer layer. Results show that the number of rollers per streamwise extent of the computational domain can be controlled by setting different the resonant frequencies; higher resonant frequencies create smaller rollers with higher convection

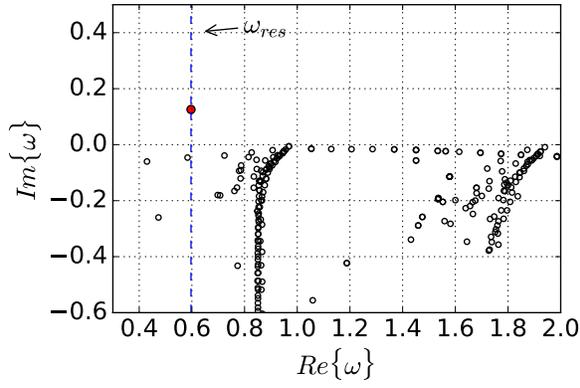
(a)

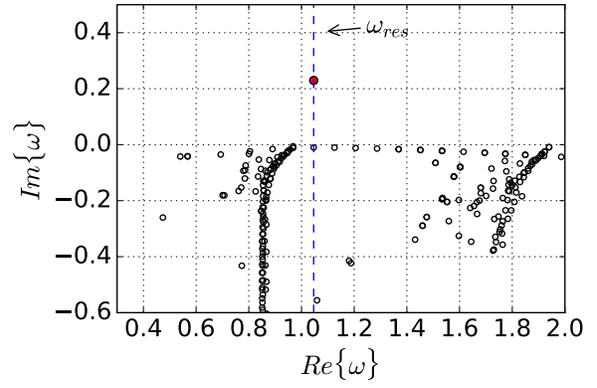
(b)

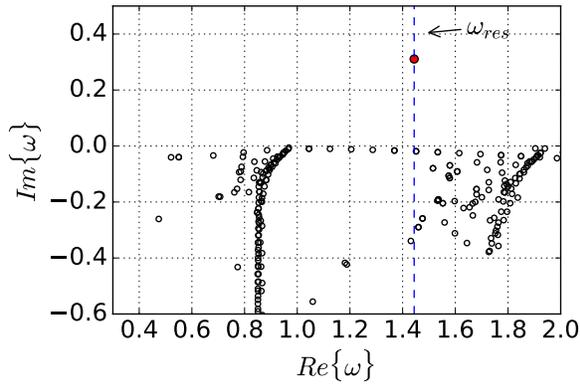
(c)

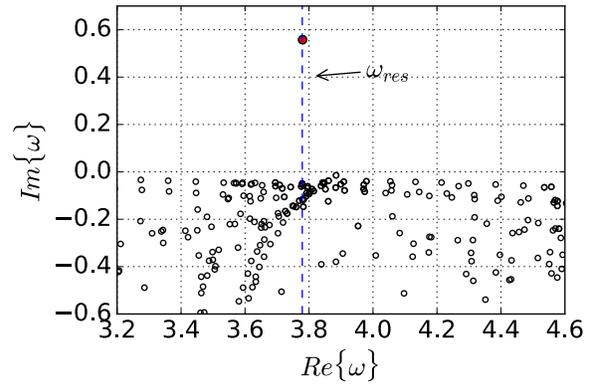
(d)

Figure 7. **Eigenvalue spectrum at** $M_b$ = 0.85 **and** $Re_b$ = 6900 **on the grid** 128 × 128 **for flow over complex impedance** (2) **at four different resonant frequencies. (a):** $\omega_{res}$ = 0.596, **(b):** $\omega_{res}$ = 1.046, **(c):** $\omega_{res}$ = 1.443, **(d):** $\omega_{res}$ = 3.779

velocities. The extent of the rollers into the outer layer can also be controlled by varying this parameter. Smaller rollers that are associated with the higher frequencies have less impact on the outer layer since their magnitude is more rapidly evanescent in the wall-normal direction.

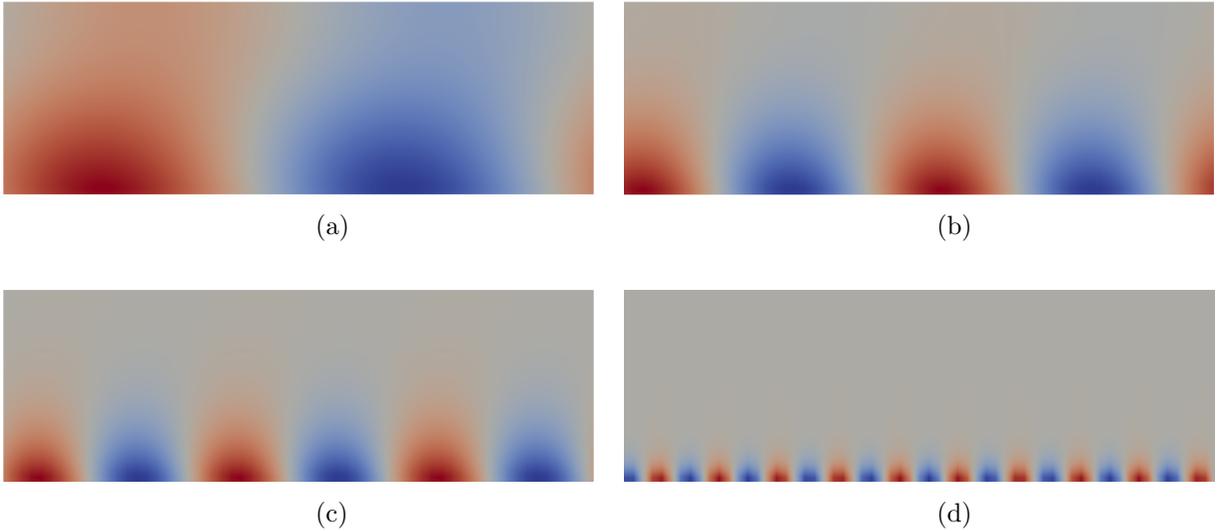

Figure 8. Pressure eigenfunction of the unstable modes at $M_b = 0.85$ and $Re_b = 6900$ on the grid $128 \times 128$ for four different resonant frequencies. (a): $\omega_{res} = 0.596$, (b): $\omega_{res} = 1.046$, (c): $\omega_{res} = 1.443$, (d): $\omega_{res} = 3.779$

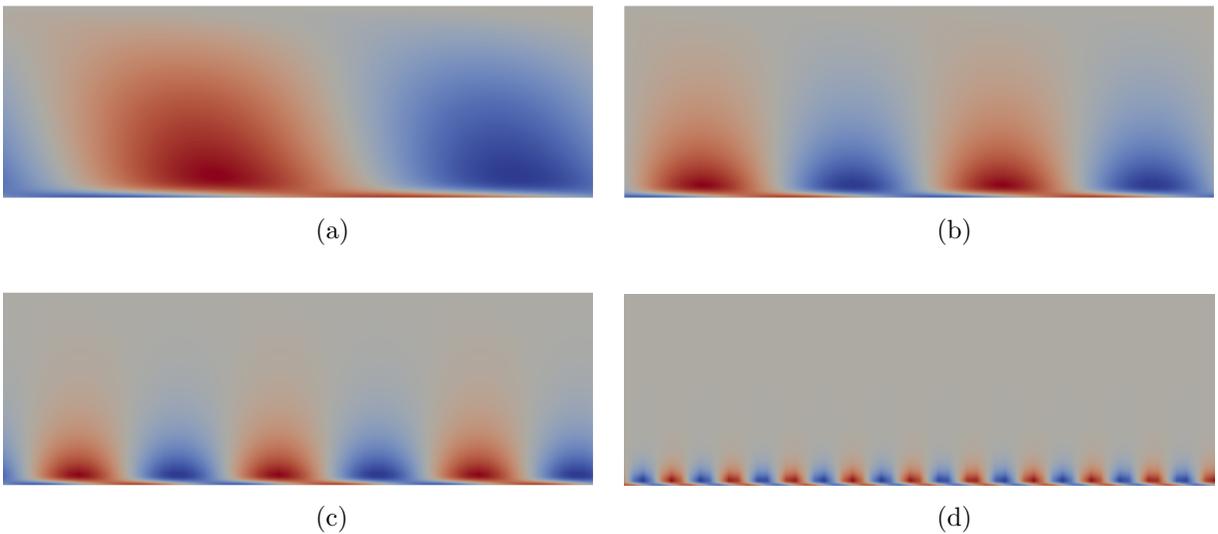

Figure 9. Wall-normal velocity eigenfunction of the unstable modes at $M_b = 0.85$ and $Re_b = 6900$ on the grid $128 \times 128$ for four different resonant frequencies. (a): $\omega_{res} = 0.596$, (b): $\omega_{res} = 1.046$, (c): $\omega_{res} = 1.443$, (d): $\omega_{res} = 3.779$

In the all calculations presented before, impedance of the bottom wall is distributed uniformly in the streamwise direction. In the final test case, we are interested in studying the effect of spatially varying resonant frequency on the turbulent structures. To this end, the resonant frequency is varied about a nominal value following a sin function:

$$\omega_{res} = \omega_0 \times (1 + 0.5 \sin(x)) \quad (25)$$

where $\omega_0$ is set to 1.443 in order to accommodate a proper range of resonant frequencies in the $x$-direction. A schematic view of the channel is plotted in figure 10 (a) which qualitatively demonstrates the structure of resonators having the variable resonant frequency by considering different chamber volumes for the cavities.

The eigenspectrum for this problem is plotted in figure 10 (b) for which the linearized Navier-Stokes equations are discretized on a 64 × 128 grid. All the other parameters are kept the same as the previous cases. In this figure, the shaded area shows the stable region while red circles refer to the first and second unstable modes. The frequency range in which the resonant frequency varies spatially in the streamwise direction is illustrated by vertical dashed lines.

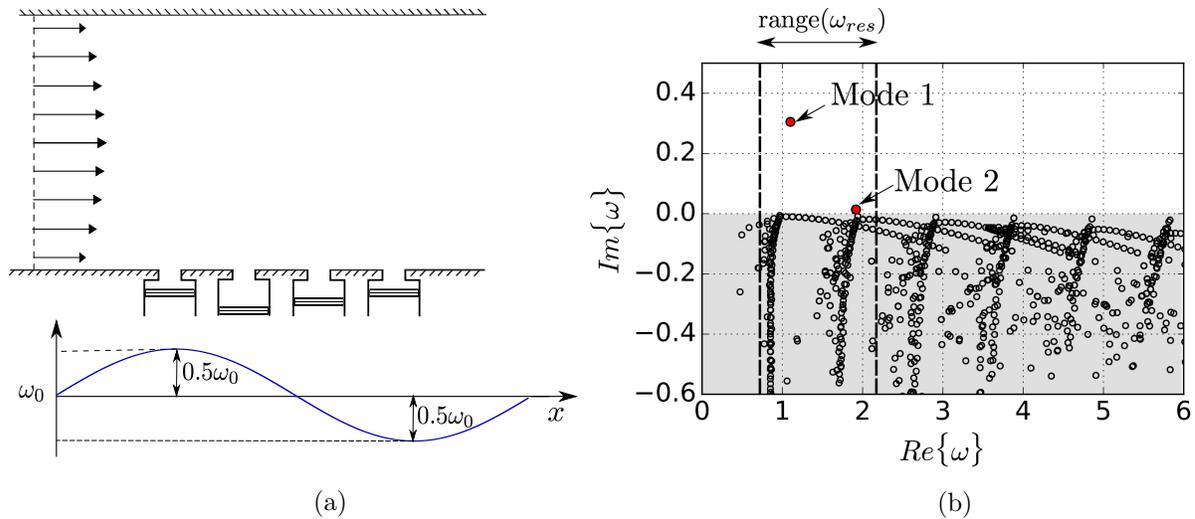

(a) (b)

Figure 10. Computational setup (a) and eigenspectrum of the channel flow with impedance bottom wall variable in streamwise direction (b). Mode 1: $\omega = 1.1005 + 0.3050j$ and mode 2: $\omega = 1.9174 + 0.0139j$

The wall-normal velocity and pressure eigenfunctions corresponding to these two modes are plotted in figure 11.

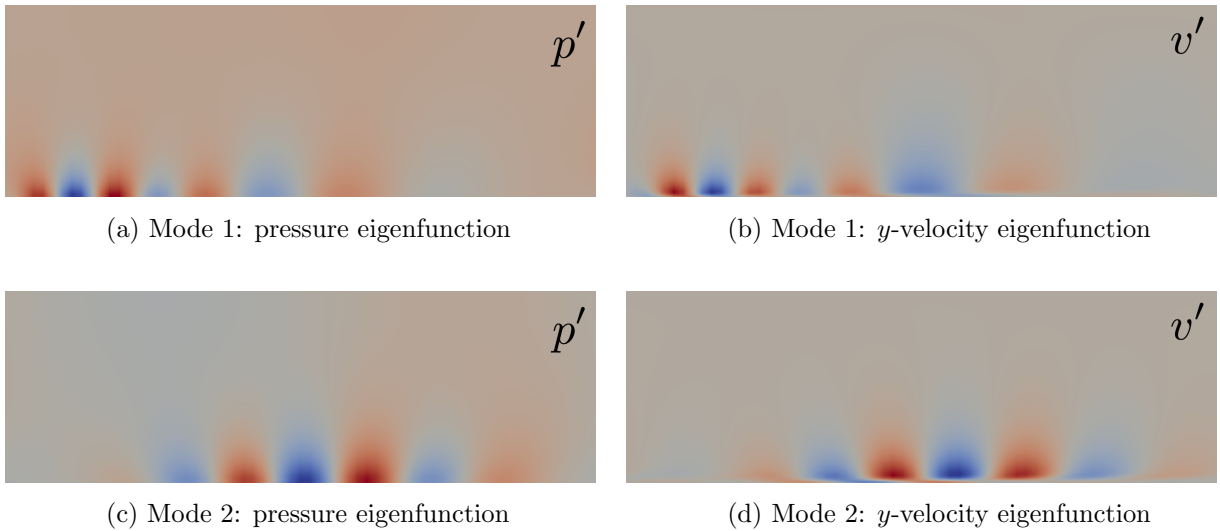

(a) Mode 1: pressure eigenfunction  (b) Mode 1: $y$-velocity eigenfunction

(c) Mode 2: pressure eigenfunction  (d) Mode 2: $y$-velocity eigenfunction

Figure 11. Pressure and wall-normal velocity eigenfunctions of the unstable modes at $M_b = 0.85$ and $Re_b = 6900$ on the grid 64 × 128 for the case of variable resonant frequency in streamwise direction. Mode 1: $\omega = 1.1005 + 0.3050j$ and mode 2: $\omega = 1.9174 + 0.0139j$

Considering the distribution proposed for the resonant frequency in (25) and results presented in the last section (figures 8 and 9), one may conclude that smaller rollers (but with more numbers) should exist in the left side of the channel, due to high resonant frequency in that region. As flow travels in $x$- direction, rollers must be elongated to comply with the lower resonant frequencies imposed at those regions so that less, but wider, rollers are expected. This is the first order effect of the impedance boundary condition following the prescribed frequency distribution which is reflected in mode 1 figure 11 (a) and (b). The second mode, shown in 11 (c) and (d) predicts high intensity rollers in the middle of domain which can be considered as the second order effect of such impedance boundary condition. Both modes suggest that the bottom wall turbulence structure does not affect the ones at the top wall.

The existence of these modes are supported by the preliminary results of numerical simulation of fully non-linear Navier-Stokes Equations for the prescribed computational setup using the above-mentioned Navier-Stokes solver, CFDSU. The computational domain considered for this simulations is $L_x \times L_y \times L_z = 2\pi \times 2 \times 1.5\pi$ which is discretized on a grid of size $N_x \times N_y \times N_z = 96 \times 196 \times 96$ resulting in a coarse DNS resolution.

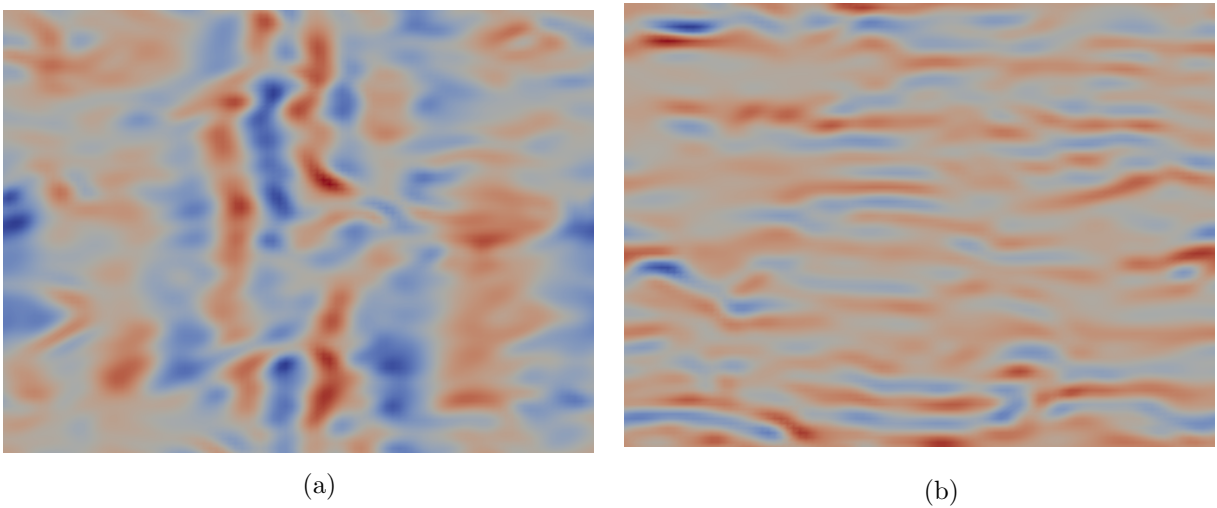

(a)          (b)

**Figure 12.** Wall normal velocity contour on the bottom (a) and top wall (b) for a channel flow with impedance boundary condition on the bottom wall wherein the resonant frequency varies sinusoidally. Domain size is $L_x \times L_y \times L_z = 2\pi \times 2 \times 1.5\pi$ which is discretized with a grid of size $N_x \times N_y \times N_z = 96 \times 196 \times 96$

Figure 12 (a) which corresponds to the bottom wall, clearly shows the elongated rollers at the right side of the channel. The small rollers at the far left of the channel can also be seen in the time-varying snapshots, however, is not well reflected in the instantaneous visualization presented here. The presence of the high intensity small rollers in the middle of the channel supports the physical likelihood of the second mode. On the contrary, the wall normal velocity contour at the top wall, shown in figure 12 (b) reveals the classic near wall turbulent streaks suggesting that the alteration of turbulent structures at the bottom wall does not significantly affect the flow near the top wall which has also been concluded using the results of the global stability analysis.

## Acknowledgments

The authors have also significantly benefited from a generous computational allocation on Purdue's eighth on-campus supercomputer, *Rice*.